# Ballistic transport experiment detects Fermi surface anisotropy of graphene

Takushi Oka[1], Shingo Tajima[1], Ryoya Ebisuoka[1], Taiki Hirahara[1], Kenji Watanabe[2], Takashi Taniguchi[2,] and Ryuta Yagi[1*,]

1 Graduate School of Advanced Sciences of Matter (AdSM,) Hiroshima University, Kagamiyama 1-3-1, Higashi-Hiroshima 739-8530, Japan,
2 National Institute for Materials Science (NIMS), Namiki 1-1, Tsukuba 305-0044, Japan.

ABSTRACT

Monolayer graphene and bilayer graphene have strikingly different properties. One such difference is the shape of the Fermi surface. Although anisotropic band structures can be detected in optical measurements, they have so far been difficult to detect in transport experiments on two-dimensional materials. Here we describe a ballistic transport experiment using high-quality graphene that revealed Fermi surface anisotropy in the magnetoresistance. The shape of the Fermi surface is closely related with the cyclotron orbit in real space. Electron trajectories in samples with triangular lattices of holes depend on the anisotropy of the Fermi surface. We found that this results in the magnetoresistance which are dependent on crystallographic orientation of the antidot lattice, which indicates the anisotropic Fermi surface of bilayer graphene which is a trigonally-warped circle in shape. While in monolayer, shape of magnetoresistance was approximately independent of the orientation of antidot lattice, which indicates that the Fermi surface is a circle in shape. The ballistic transport experiment is a new method of detecting anisotropic electronic band structures in two-dimensional electron systems.

## 1. Introduction

Electronic band structures of monolayer graphene and bilayer graphene are strikingly different [1], and the difference appears in various properties. One example is the bandgap between the conduction and valence band. Monolayer graphene is a semi-metal and does not have a band gap. However in bilayer graphene, a band gap appears by applying a perpendicular electric field [2-5]. The difference in the band structure can be directly revealed in the Landau level structures in the measured magnetoresistance. The difference between mono- and bilayer graphene lies in the shape of the Fermi surfaces as well. Fermi surfaces of graphene are more or less trigonally-warped circles; *i.e.*, the Fermi surfaces are deformed circles with three-fold rotational symmetry. However, trigonal warping of monolayer graphene is negligible in the low-energy regime, $E < 250$ meV, where samples used in transport experiments typically have carrier densities of $|n| < 4 \times 10^{12}$ cm$^{-2}$. On the other hand, band calculations indicate that the Fermi surface of bilayer graphene is a rounded triangle (see Fig. 1(a)) excepting in the vicinity of the charge neutrality point. The shape of the Fermi surface is closely related with the shape of the electron orbit in real space in the presence of a magnetic field. In the semi-classical picture, the wave vector of an electron varies according to semi-classical equation of motion under the influence of Lorentz force. The wave vector evolves along the Fermi surface. If the Fermi surface is circular, the real space orbit of the electron becomes a circle, and if the Fermi surface is a deformed circle, the real space orbit of the electron is a deformed circle (see Fig. 1 (a)). This shape of orbit would be able to be detected by measuring magnetoresistance due to ballistic transport of electron in magnetic field in samples with electron reflectors that interfere with electron cyclotron motion. An example of such samples is triangular graphene antidot lattices with holes drilled in the graphene to form a triangular lattice.

The electron trajectory in the antidot lattice would vary significantly if the cyclotron orbit is a circle or deformed circle because of the reflection at the antidots. The difference between electron trajectories due to reflections can be detected by using the commensurability magneto- resistance of the antidot lattices. If the density of impurity in the conductor is sufficiently low, the electron mean free path ($l_f$) becomes sufficiently long, and electron's motion in the conductor becomes ballistic and its orbit becomes a circle in magnetic field. Collisions with antidots generally complicates trajectory of the electrons. However, at the magnetic field where cyclotron diameter $R_c$ satisfies the condition,

$$2R_c = a, \qquad (1)$$

a peak appears in the magnetoresistance that is due to the commensurability of the lattice period and the cyclotron diameter [6-10]. Here, $a$ is the distance between centers of neighboring antidots. This is condition 1 illustrated in Fig. 1(b), where an electron starting from an antidot collides with the next nearest antidot. More resonant conditions are possible, *e.g.*, conditions 2 and 3 in lower magnetic fields.

That the cyclotron orbit from one antidot to the other affects magnetoresistance properties suggests that the antidot lattice could be used to detect anisotropy of the Fermi surface [11]. To illustrate the basic idea, here we consider the simplest cases in panels (i)-(iii) of Fig. 1 (b). Panel (i) is for an isotropic Fermi surface. Rotating the antidot lattice would not result in a significant change in electron trajectories, so that the magnetoresistance would be unchanged. On the other hand, in the case of a trigonally warped Fermi surface, rotation of the antidot lattice results in a significant difference in the electron trajectory from one antidot to another, as shown in panels (ii) and (iii). In panel (ii), long partial cyclotron orbits from an antidot to the other are possible as shown by the arrows. However, no such long trajectories are possible in panel (iii). The rotation of the antidot lattice results in a difference in the possible partial cyclotron orbits and would result in a different magnetoresistance. This kind of Fermi surface anisotropy has not been reported in the past studies on graphene antidots [12-14].

## 2. Simulation based on Model Fermi surface

Before we show our experimental results, we verify the above expectations through numerical simulations. We performed a numerical calculation of magnetoresistance in a triangular antidot lattice. The conductivity of this system in the presence and in the absence of magnetic fields can be calculated from [15],

$$\sigma_{ij} = A \int_0^\infty < v_i(0)v_j(t) >_{av} e^{-t/\tau} dt. \qquad (2)$$

Here, $v_i$ and $v_j$ are the $i$ and $j$ components of the group velocity of the wave packet and are calculated by using the semi-classical equation of motion. $< \cdots >_{av}$ is an average over all possible initial states in phase space, $\tau$ is the relaxation time associated with impurity scattering, and $A$ is a constant. The conductivity calculated with equation (2) is strongly influenced by deformation of the Fermi surface. In the actual calculation we evaluated eq. (2) using a model Fermi surface that has an analytic form,

$$k = k_0(1 + \alpha \cdot \cos(3\theta)). \qquad (3)$$

Here $(k, \theta)$ denotes polar coordinates describing the magnitude of the wave vector and azimuthal angle. The parameter $\alpha$ tunes the degree of trigonal warping. When $\alpha = 0$, the Fermi surface is circular and isotropic. A slight trigonal warping in monolayer graphene can be expressed by $\alpha \approx 0.01$. The resultant shape of Fermi surface is virtually unchanged from the case of $\alpha = 0$. The model Fermi surface with $\alpha = 0.1$ is approximately that of bilayer graphene calculated by a band calculation, for carrier densities at which experiments are usually done (see Appendix 2).

Electron (or hole) trajectories were calculated semi-classically assuming that the electrons (or holes) are specularly reflected at the boundary of the antidots. The antidot lattice can be characterized by $d/a$, where $a$ is the distance between the center of adjacent antidots and $d$ is the diameter of the antidot. In the calculation, $d/a = 0.2$ was used.

Figure 2a shows results for $\alpha = 0.01$. The horizontal axis $a/R_c = eaB/(\hbar k_0)$ is proportional to the magnetic field. This is the case for monolayer graphene which has an approximately circular Fermi surface. The magnetoresistance does not significantly vary with $\theta$. Oscillatory peaks are visible at $l_f > a$. The highest peak, marked with an arrow at $a/R_c = 2$ is the commensurability peak associated with the nearest neighbor antidots (the case 1 in Fig. 1(b)).

The shape of the background magnetoresistance (low field magnetoresistance without peaks for commensurability) is relevant to the degree of the anisotropy of the Fermi surface. Figures. 2 (b)-(d) show the results of the numerical simulation for different values of $\alpha$. It is clear that, with increasing anisotropy parameter α, positive background magnetoresistance appears in low magnetic fields, and it show a significant $\theta$-dependence. In addition, for small α ( $\leq 0.2$ ), magnetic fields for commensurability peaks are approximately unchanged while they show variations for larger α.

## 3 Experimental Results

In order to detect Fermi surface anisotropy, we performed an experiment on samples of antidot lattices with primitive vectors having different directions as shown in Fig 3(a). Fermi surface anisotropy can be detected through the relative orientation of the crystal lattice and reciprocal lattice. In the effective mass approximation, the $x$ –axis is often used as a zigzag direction, and the $y$-axis is the armchair direction. The directions of the $k_x$ and $k_y$ axes in reciprocal space are the same as those of the $x$ and $y$ axes in real space so that orientation of the Fermi surface can be relatively determined by the zigzag or armchair directions (see Fig. 1(a)). Mechanically exfoliated graphene flakes often have straight sample edges. These edges are presumably zigzag or arm chair types [16-18]. We fabricated two set of antidot lattices with the same lattice constant but with different orientations on the same graphene flake (Fig. 3 (a) and (b)). One of the antidot lattices has a primitive vector parallel to the cleaved line (possibly a zigzag edge). The other has a primitive vector tilted by 30 degrees relative to the cleaved line.  The distance between the centers of the antidot was 700 nm and diameter of the antidots was about 200 nm.

Our high-quality graphene samples is encapsulated by high-quality *h*-BN flakes. An optical micrograph of a sample is displayed in Fig. 3 (b). The mobility $\mu = \sigma / ne$ was estimated to be about μ = 60,000 $cm^2/Vs$ at large carrier densities. The mean free path of graphene was about a few µm, which was larger than the lattice constant of the antidot.

The magnetoresistance results for the bilayer graphene antidot samples with $\theta = 0°$ and $\theta = 30°$ showed qualitative different behavior. First, we checked that the sample is bilayer graphene. In high magnetic field, Shubnikov-de Haas oscillation was observed as shown in top and middle panel in Fig. 4(a), which is a map of the derivative of the longitudinal resistivity with respect to magnetic field $(dR_{xx}/dB)$ measured at $T$= 4.2 K, as a function of gate voltages and magnetic field. $\theta = 0°$ and $30°$ denote experimental results for antidot lattice samples whose angle between crystal axes of graphene and antidot lattice are 0 and 30 degrees, respectively. The fan-shaped structure appearing above about 1T originates from the Landau levels of bilayer graphene. The zero-mode Landau level appears near $V_g = 0$ V. It has twice the carrier density compared with other Landau levels, as can be seen in the positions of energy gaps between the Landau levels shown by bars at the top of the figure. This is consistent with the fact that the zero-mode Landau level of bilayer graphene has eight-fold degeneracy, whereas the other Landau levels have four-fold degeneracy. A plot of Landau level index $N$ vs. $1/B$ shows that an extrapolation of the linear relation to $1/B = 0$ gives an intersection of $N = 0$ as shown in the bottom panel in Fig 4(a), which indicates bilayer graphene [1].

The commensurability peak associated with the nearest neighbor antidots is at about $B = 0.6$ T, which is marked with the arrows in Fig. 4 (a). In the bilayer case, the commensurability condition is approximately given by Eq. (1). Accordingly the magnetic field $B_p$ for the peak is given by $B_p = 2\hbar\sqrt{\pi n}/(ea)$; i.e., $B_p$ shows a square-root dependence on carrier density because $n \propto V_g$ [13]. The square root dependence on $n$ of the commensurability peaks is thus a sublinear, which is apparently distinct from the linear dependence of the stripes due to Shubnikov-de Haas oscillations appearing at higher magnetic fields.

Figure 4 (b) shows magnetoresistance traces of the samples with $\theta = 0°$ and $\theta = 30°$ for different gate voltages. In both cases, we can see some peaks below $B < 1$ T. The largest peaks indicated by the arrows are commensurability peaks arising from matching of the cyclotron diameters with the distance between the centers of the nearest-neighbor antidots. The peaks obey the condition, $B_p = 2\hbar\sqrt{\pi n}/(ea)$. Peaks that appear in lower magnetic fields are due to a similar commensurability effect relevant to the next-nearest-neighbor antidots, the second-nearest-neighbor-antidots (cases 2 and 3 in Fig. 1 (b)), *etc.* Oscillations appearing between $B = 1$ and 2 T are due to Shubnikov-de Haas effect. The most important difference between these figures is in the overall shapes of the low field magnetoresistance without peaks for the commensurability (back ground magnetoresistance). In particular, the sample with $\theta = 0°$ shows positive background magnetoresistance in low magnetic field, while samples with $\theta = 30°$ shows slightly negative background magnetoresistance. The observed magnetoresistance is not due to Hall resistivity arising from the sample geometry. Any contribution from the Hall resistivity to the data is removed by averaging the magnetoresistance traces for $B$ and $-B$. Data in the vicinity of the charge neutrality point ($V_g = 0$ V) showed different behaviors from that of the simulation. Magnetoresistance is affected by quantization of Landau levels with small Landau indices and by the divergent magnetoresistance reported in, for example, Refs. [19, 20].

On the other hand, significant anisotropy was not observed in the monolayer graphene antidot lattice samples. Results are displayed in Fig. 4 (c) and (d). As in the case of bilayer graphene, a fan-shaped structure is discernible in the mapping plot of $dR_{xx}/dB$ with respect to $V_g$ and $B$. From this fan diagram, one can verify that the measured sample is single-layer graphene, because the zero-mode Landau level is four-fold degenerated, as can be seen from the identical intervals between the adjacent gaps at high magnetic fields. A plot of index of Landau levels $N$ *vs* $1/B$ showed the intersection of $N = 0.5$ at $1/B = 0$ as shown in the bottom panel in Fig. 4(c), which indicates the Berry phase of π in monolayer graphene. Commensurability magnetoresistance peaks appear at the magnetic fields indicated by the arrows in the figure. There is no significant difference between the results of the samples with $\theta = 0°$ and $\theta = 30°$, and this contrasts with the results for bilayer graphene.

An important feature of our experimental data on bilayer graphene is approximately reproduced by the simulation with α = 0.1, as shown in Fig. 2 (b). It is clear that the magnetoresistance traces show significant $\theta$ dependence. In particular, the results for $\theta = 0°$ show a positive background magnetoresistance as compared with the result for $\theta = 30°$, which even shows slightly negative background magnetoresistance, is similar to the monolayer case. This indicates that the model Fermi surface with $\alpha = 0.1$ approximates the actual Fermi surface of bilayer graphene for the carrier density regime of the experiment, $1 - 3 \times 10^{12}$ cm$^{-2}$. This is consistent with results of the band calculation (see Appendix 2). Shubnikov-de Haas oscillations, which were observed above $B$ = 1 T, do not appear in our simulation based on the semi-classical model.

4. Discussion

Graphene nanoribbons are another kind of system in which the crystal axis plays a crucial role, as in our system. The electronic band structure is predicted to be highly dependent on the edge structure (zigzag or armchair) of the nanoribbon [21-25]. To date, there have been experiments on commensurability magnetoresistance in monolayer graphene, where the rotation angle of crystal lattice relative to the antidot lattice was not considered [13, 14]. The present result is consistent with the previous findings. On the other hand, we previously reported on commensurability magnetoresistance in monolayer and bilayer graphene antidot samples where rotation angles of crystal lattice and antidot lattice were not determined clearly [13]. In the study we found that the shape of the magnetoresistance of bilayer graphene antidots was qualitatively different from that of monolayer graphene, but the origin of the difference remained unclear. Here we have shown that it is possibly due to Fermi surface anisotropy.

Graphene nanoribbons are another kind of system in which the crystal axis plays a crucial role, as in our system. In principle, it would be possible to detect the edge structure, zigzag or armchair, by measuring the transport properties of the nanoribbon, and determine the crystallographic orientation of the edge. However, at the present stage, the roughness of the edge poses a significant problem to making transport experiments [26-36]. On the other hand, one can extract information on the orientation of the crystallographic axes by using antidot lattices. Because we measured ballistic electron transport, issues regarding edge roughness are virtually irrelevant.

A magneto-focusing experiment [37] is similar to the antidot lattice experiment. In a local picture, the antidot experiment is approximately the same as a transverse electron magneto-focusing experiment [37]. An important difference is the presence of many antidots in the sample. They serve as reflectors during the magneto-focusing-like process from one antidot to the other antidots. The electron trajectory is determined by the magnetic field, antidot lattice parameters, and the shape of the cyclotron orbit which reflects the shape of the Fermi surface, and it results in a different magnetoresistance. It is not clear whether Fermi surface anisotropy can be detected by using the magneto-focusing effect. We expect that it can be observed by scanning gate microscopy to visualize the spatial distribution of the cyclotron motion [38].

Regarding the methods of detecting the shape of Fermi surface through transport measurements, Shubnikov-de Haas oscillations can be measured for various magnetic field angles in three dimensional samples [39]. Moreover, in the case of cylindrical Fermi surfaces, angular-dependent magnetoresistance oscillations (AMROs) can be used to map out shape of Fermi surface [40-43]. However, in two-dimensional electron systems, a tilted magnetic field cannot be used get information on the shape of the Fermi surface. Ballistic transport experiments using antidots are thus promising means of probing the Fermi surfaces of a variety of two-dimensional materials.

Summary

Commensurability magnetoresistance in antidot lattices reflects shape of the cyclotron orbit, which again reflects the shape of the Fermi surface. We found that this results in background magnetoresistance (low field magnetoresistance without peaks for commensurability) that is strongly dependent on the crystallographic orientation of graphene and antidot lattice. By conducting magneto-transport measurements using antidot lattice, we have demonstrated anisotropic Fermi surface in bilayer graphene and approximately isotropic Fermi surface in monolayer graphene. Observed behavior was explained by calculations within semi-classical theory. This method can be used to detect Fermi surface anisotropy in two-dimensional materials.

Appendix

1 Sample fabrication and magnetotransport measurements

Our graphene antidot lattice samples were made using encapsulated graphene formed by using the technique described in Ref. [44]. Figure 5 shows the steps of sample fabrication. An exfoliated graphene was picked up with a thin *h*-BN flake formed on PPC (poly (propylene carbonate)) film by using mechanical exfoliation of bulk *h*-BN crystal. Then the *h*-BN-graphene stack was transferred onto an *h*-BN flake on $SiO_2$/Si substrate to form encapsulated graphene. Heavily doped Si substrate, which is conducting even at low temperature, served as a back gate. An antidot lattice was formed by using standard electron beam lithography. Organic electron beam resist was coated on the encapsulated graphene, and it was patterned into triangular lattices of small holes. Then the encapsulated graphene it was plasma-etched with a mixture of low pressure $CF_4$ and $O_2$ gas to form an antidot lattice structure.

Electric contacts were formed by using the technique described in [44]. Electric leads were formed by using electron beam lithography and the lift off technique.

The magnetoresistance of the samples was measured at $T$= 4.2 K by applying a perpendicular magnetic field with a superconducting solenoid. Resistance was measured by using the standard lock-in technique.

2 Model Fermi surface

Figure 6 **a** shows a constant energy contour plot of the dispersion relation for mono- (right) and bilayer (left) graphene. The dispersion relation was calculated on the basis of the effective mass approximation. Slonczewski-Weiss-McClure parameters were the same as those of graphite, $\gamma_0 = 3.16$ eV, $\gamma_1 = 0.39$ eV, $\gamma_2 = -0.02$ eV, $\gamma_3= 0.3$ eV and $\Delta_p = 0.037$ eV. The energy for the contour plot is for carrier densities of $|n|< 4 \times 10^{-12}$ cm$^{-2}$. The Fermi surface of the monolayer graphene is approximately circular while that of the bilayer is a rounded triangle. To calculate the magnetoconductivity

component, we used a model two-dimensional Fermi surface $(k_x, k_y)$ which is described by the polar axis $(k, \phi)$ as,

$$k = k_0(1 + \alpha \cos(3\phi + \phi_0)). \tag{4}$$

Here, α is a parameter that describes trigonal warping. $\phi_0$ is a parameter that specify valleys, *i.e.*, K or K' point in the reciprocal space. $\phi_0 = 0$, and π represents K and K' valleys, respectively. Figure 6(b) shows the shapes of the model Fermi surfaces for different values of α. Monolayer graphene is for $\alpha = 0$, and the Fermi surface is a circle. The result with $\alpha = 0.1$ approximates the shape of the Fermi surface of bilayer graphene.

## 3 Trigonal waring of monolayer graphene.

Low energy band structure of monolayer graphene near K and K' points can be described by the Hamiltonian based on the effective mass approximation [45].

$$\hat{H} = \begin{pmatrix} 0 & v_0(\xi\hat{p}_x - i\hat{p}_y) \\ v_0(\xi\hat{p}_x + i\hat{p}_y) & 0 \end{pmatrix}. \tag{5}$$

Here, $v_0 = \sqrt{3}a\gamma_0/2\hbar$, and $a$ is the lattice constant of graphene. $\hat{p}_x$ and $\hat{p}_y$ are momentum operators, respectively. ξ is a valey index for K ($\xi = 1$) and K' ($\xi = -1$) valleys. This Hamiltonian leads to dispersion relations which is isotropic, and hence the Fermi surface is circular. Trigonal warping of monolayer graphene originates from a higher order term in the *kp* scheme [46,47]. We have numerically calculated the dispersion relation considering the additional term (the equation (3.6) in Ref. [46] with $\theta = 0$) using the Slonczewski-Weiss-McClure parameter of graphite ($\gamma_0 = 3.16$ eV, $\gamma_1 = 0.39$ eV). The energy-contour is shown in Fig. 7 (a). For large energies ($E = \pm$ 1eV), trigonal warping similar to that of bilayer graphene is clearly visible. With decreasing $|E|$, the effect of trigonal warping tends to vanish. Fig. 7 (b) is the energy contour at $E = \pm$ 0.25 eV, at which the carrier density is expected to be about $4 \times 10^{12}$ cm$^{-2}$, which is larger than the maximum carrier density in the present transport experiments. The red solid line is for results of calculation by considering trigonal warping. This approximately coincides with a circular one (blue dashed line) which is calculated without considering trigonal warping.

4 **Scatterings and magnetoresistance.**

How an electron wave packet is reflected at the graphene edge is an important problem in graphene research. It is still unclear experimentally because of the difficulty of forming graphene samples that have particular edge inclinations. Taychapatanat *et al.*, by observing the magneto-focusing effect with higher order peaks, found that the reflections of electron by the sample edge are specular [37]. Masubuchi *et al.* reported that the reflections are diffusive, by performing edge scattering experiments using nano-wire [48]. The edge structure of our antidot samples would be complicated, presumably very rough, rather than zigzag or armchair because they were made by drilling holes with a plasma etching process. Because the nature of the reflections at rough edges is unknown, we studied two extreme cases. One was specular reflection. The result of the simulation is displayed in left panel in Fig. 8. Here, we assumed the antidots to be circles with the same diameter, and electrons are reflected specularly at their boundary. The other case is diffusive reflection, where electrons are reflected in a random direction. We assumed that electrons are uniformly reflected to $-(1-\delta) \times \pi \leq \theta' \leq (1-\delta)\pi$, where $\theta'$ is the angle measured from the direction of the normal vector of the antidot, and $\delta$ is a small number that was introduced for convenience, and was chosen to be $\delta = 0.05$. The right panel in Fig. 8 shows result for $\alpha = 0.1$ and $l_f/a = 2$. The results for specular reflection for the same parameters are plotted in the left panel. The simulated magnetoresistances in the cases of the diffusive and specular scattering plots have approximately the same shape.

## Acknowledgements

This work was supported by KAKENHI No.25107003 from MEXT Japan.

Figure Captions

FIG. 1 (a). Dispersion relation of graphene (i), shapes of Fermi surface (ii), and cyclotron orbit (iii) are displayed for bilayer (2L) and monolayer (1L) graphene. K and K' indicate the valley. (b). (Left) Cyclotron orbits showing commensurability peaks in a triangular antidot lattice. (Panels i-iii) Shapes of electron trajectories in magnetic field and antidot lattice.

FIG. 2 Numerical simulation of magnetoresistance of antidot lattices. Magnetoresistance calculated for different values of α. Inset in upper part of each panel shows the shape of the model Fermi surface. $l_f/a = 2$. Data were offset. (a) Results for $\alpha = 0.01$. The Fermi surface is approximately isotropic. Upper left inset shows results of simulation for different values of $l_f/a$. $\alpha = 0$. (b) Case of $\alpha = 0.1$. Upper right inset shows definition of rotation angle $\theta$ between honeycomb lattice of graphene and the antidot lattices. (c) Case of $\alpha = 0.2$. (d) Case of $\alpha = 0.3$.

FIG. 3 Sample structure.
(a) Schematic diagram of sample structure of graphene antidot lattice samples. (b) Optical micrograph of graphene antidot device. Thick white line in lower panel indicates a cleave line of graphene. Primitive vector of triangular antidot lattice was rotated by 0° and 30° from the cleave line. Probe numbers are indicated by 1-5.

FIG. 4 Commensurability peak in bilayer antidot lattice sample.
(a) (Top and middle) Maps of $dR_{xx}/dB$ with respect to magnetic field and gate voltage of bilayer graphene. Arrows indicate commensurability peaks. (Bottom) A plot of Landau index *vs.* $1/B$ for resistance minima of the S-dH oscillations. (b) Magnetoresistance traces for different gate voltages. Resistance was normalized by that at $B = 0$. From bottom to top $V_g$ was varied from -50 V to 40 V in 10 V steps. Data are shifted for clarity. Current was applied by using probe 1 and 5. Voltage drop between probes 2 and 3 was measured for $\theta = 0$, and that between 3 and 4 was measured for $\theta = 30°$. (c) Similar plot as panel (a) for monolayer graphene. (d) Similar plots as panel (b) for monolayer graphene.

FIG. 5 Sample Fabrication Process.
G is graphene, B is a thin *h*-BN flake, S is a SiO$_2$/Si substrate and P is a PPC film. Dashed white line indicates a cleaved line. Numbers from 1 to 5 indicate fabrication steps. Step 1 and 1': graphene and thin *h*-BN flakes were made by mechanically exfoliating bulk crystals. Step 2: graphene was picked up with thin *h*-BN flake on a PPC film. Step 3: the graphene/*h*-BN stack was transferred on another *h*-BN flake to form encapsulated graphene. Step 4: electron beam resist was pattered to form masks for the plasma etching process. Samples and antidot lattices were patterned by plasma etching. Step 5: samples were pattered by using plasma etching. Two antidot lattices with different orientations relative to the crystal axis were pattered using the same encapsulated graphene.

FIG. 6 Model fermi surface of bilayer graphene.
(a) Fermi Surface of bilayer graphene (left) and monolayer graphene (right), which was calculated by effective mass approximation. Constant energy surface of the dispersion relation were plotted. Slonczewski-Weiss-McClure (SWMcC) parameters for this calculation were the same as those of graphite.
(b) A model Fermi surface described by $k = k_0(1 + \alpha \cos 3\phi)$ is compared with band calculation. Here, results for a valley are displayed.

FIG. 7 Trigonal warping in monolayer graphene.
(a) Energy contours of the dispersion relation of monolayer graphene for different energies, which were calculated using SWMcC parameters of graphite. $\xi = 1$. (b) Comparison of Fermi surface shapes of monolayer graphene for $E = \pm 0.25$ eV. Solid and broken lines are for calculations with and without trigonal warping.

FIG. 8 Simulations with different scattering models.
 (Right) Magnetoresistance for diffusive reflection. $\alpha = 0$, $l_f/a = 2$.
 (Left) Results for specular reflection (the same as Fig 2(b) in the main text). $\alpha = 0$, $l_f/a = 2$.

Fig. 1

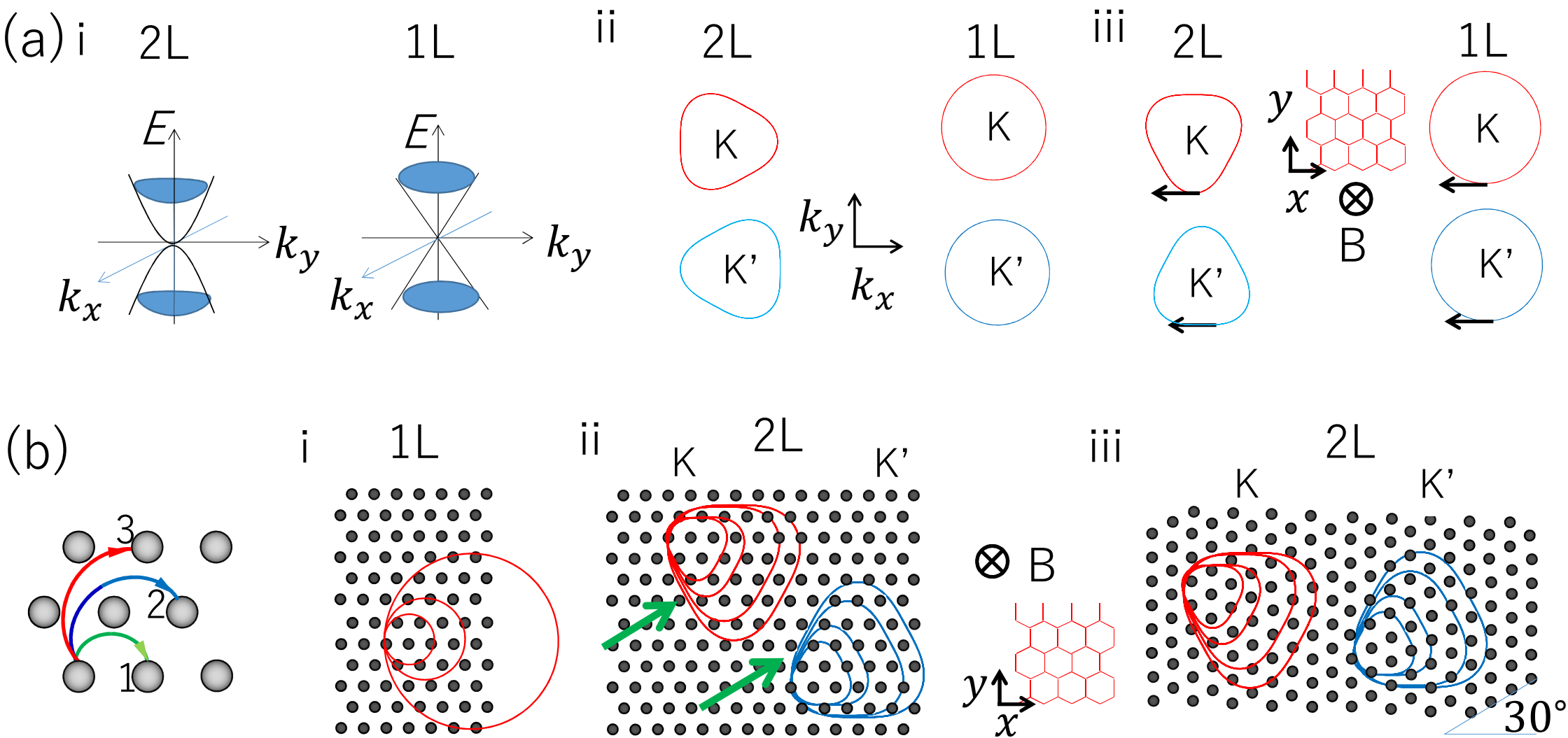

Fig2

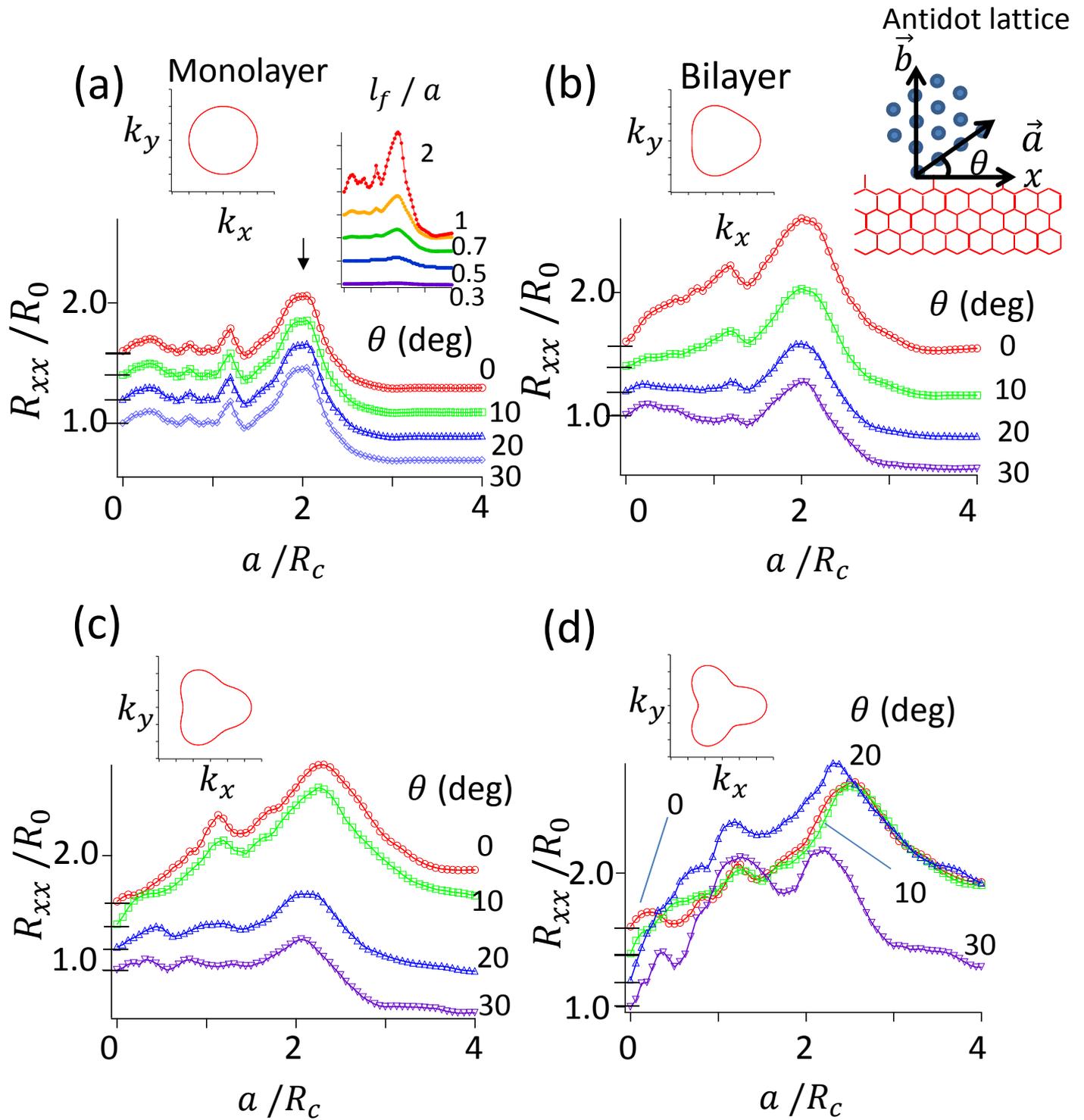

Fig3

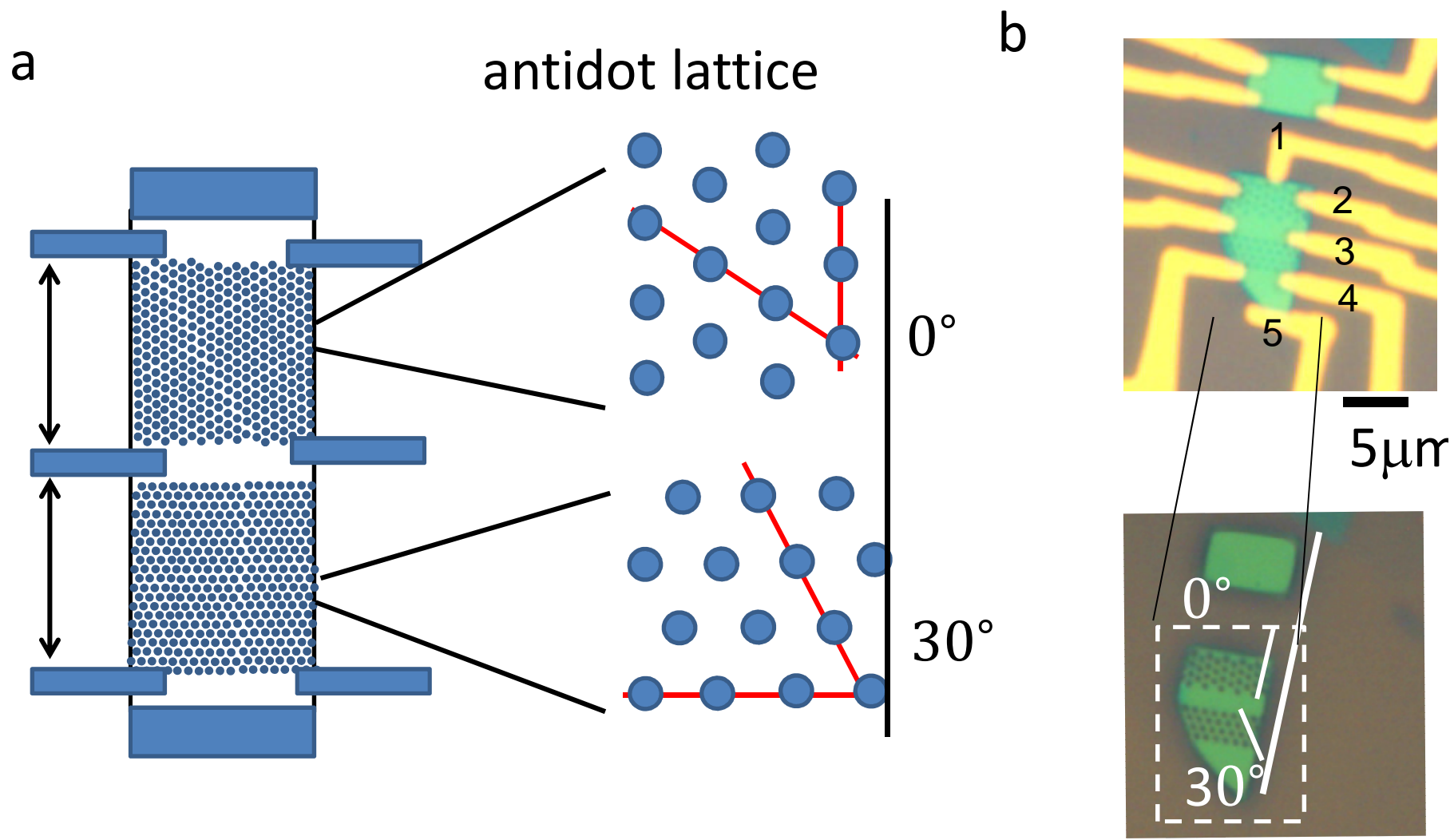

Fig. 4 (a) Bilayer (b) Bilayer

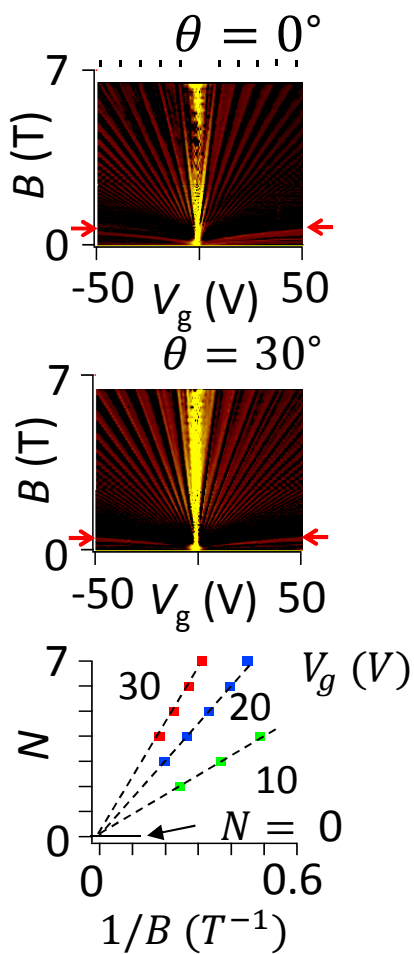
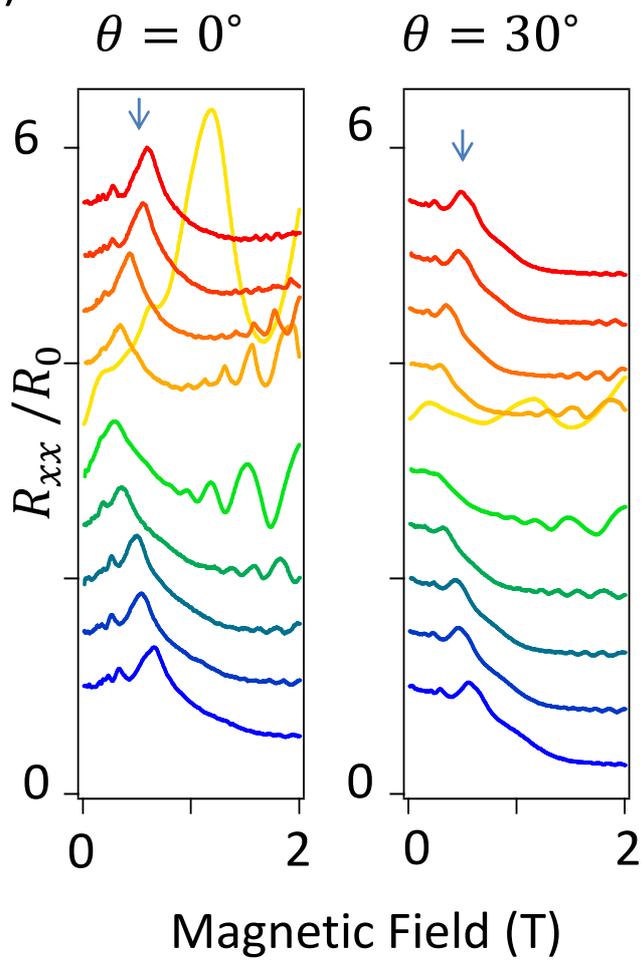

(c) Monolayer (d) Monolayer

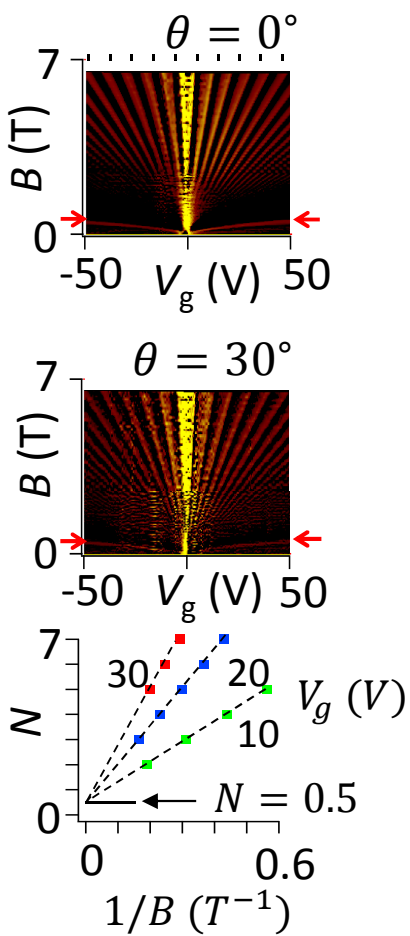
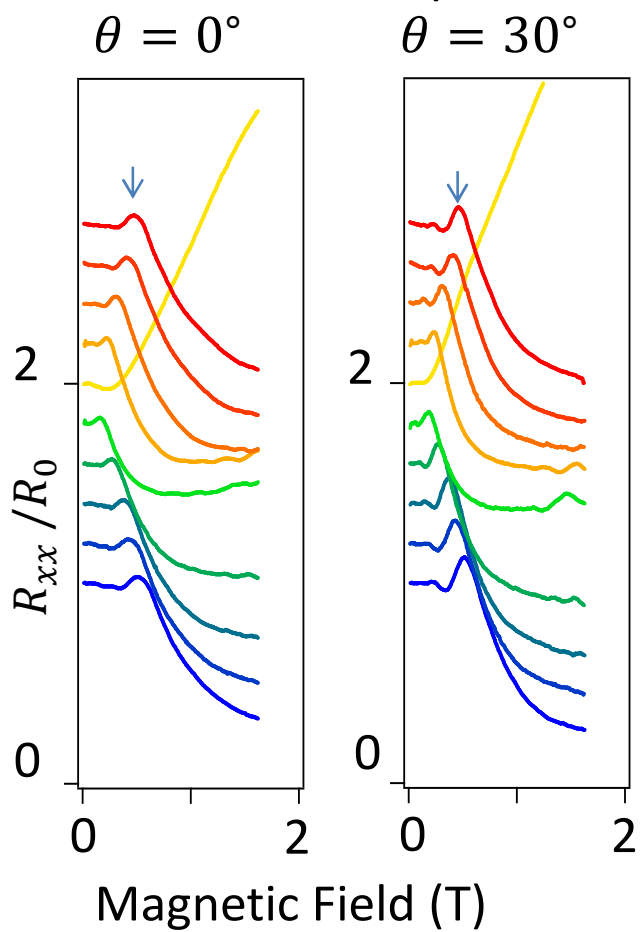

# Fig.5

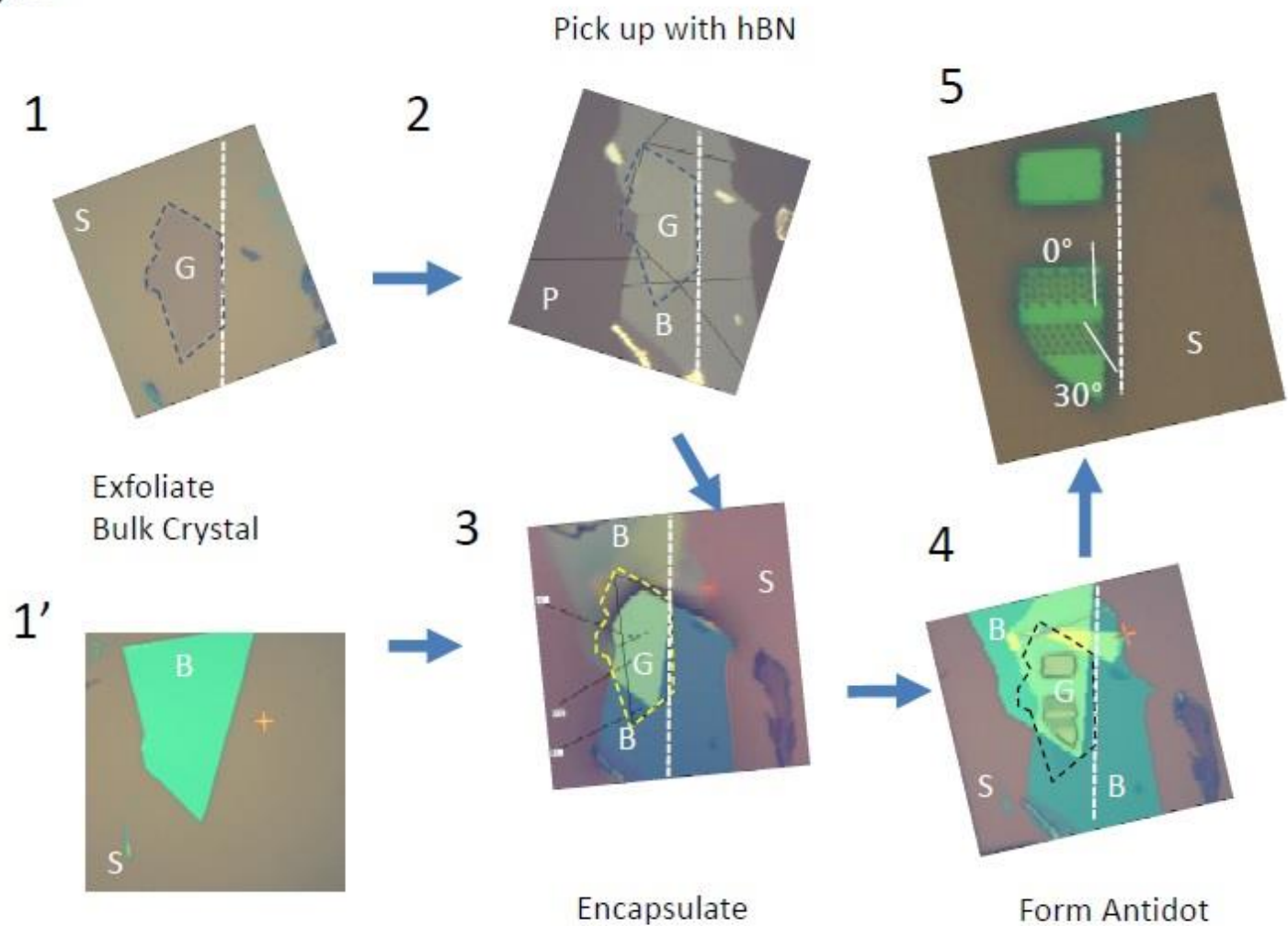

1 Exfoliate Bulk Crystal

1'

2 Pick up with hBN

3 Encapsulate

4 Form Antidot

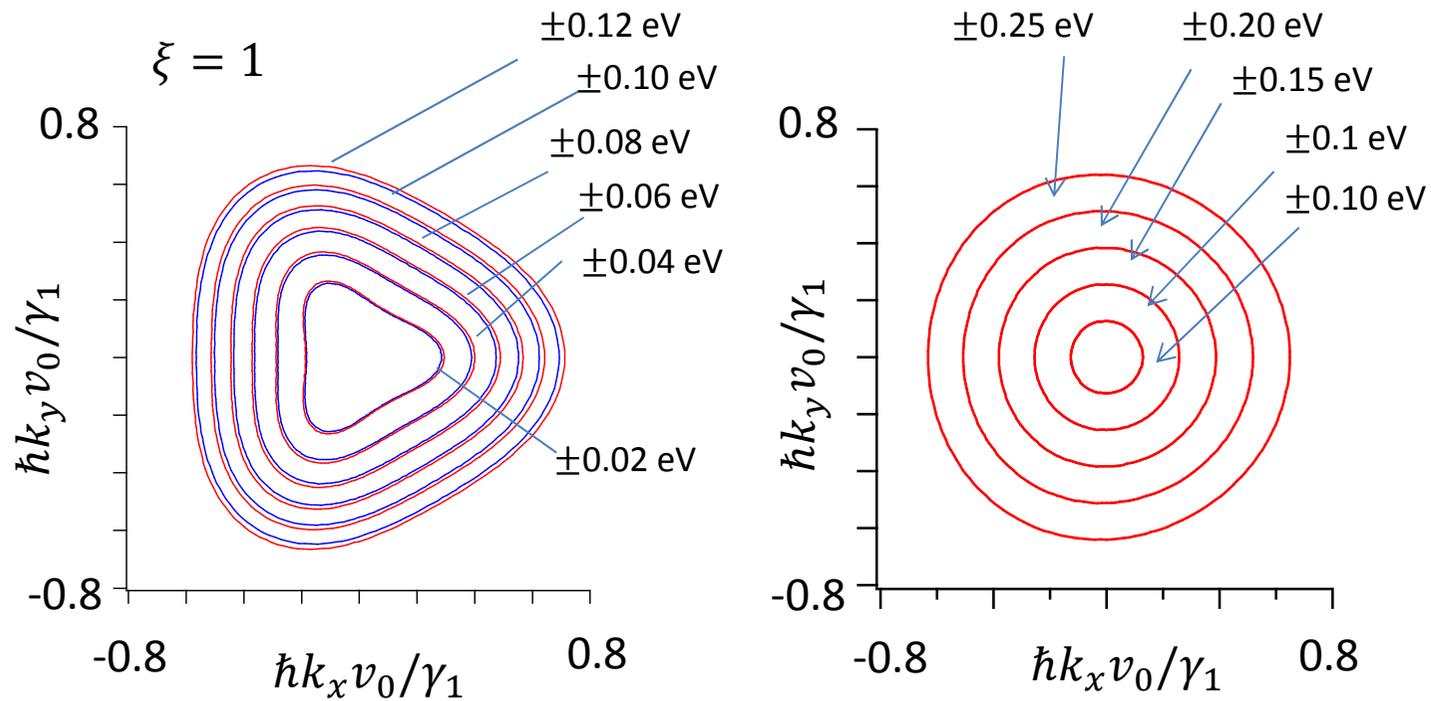
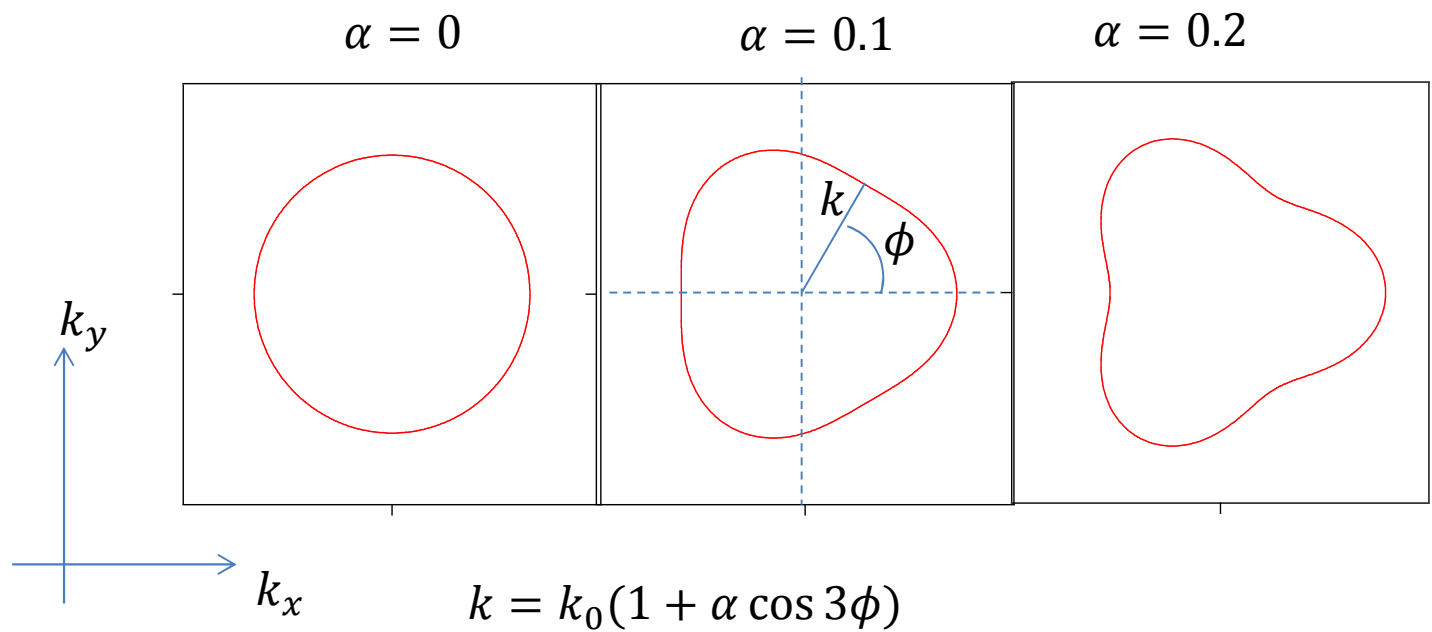

Fig. 6

Fig. 7

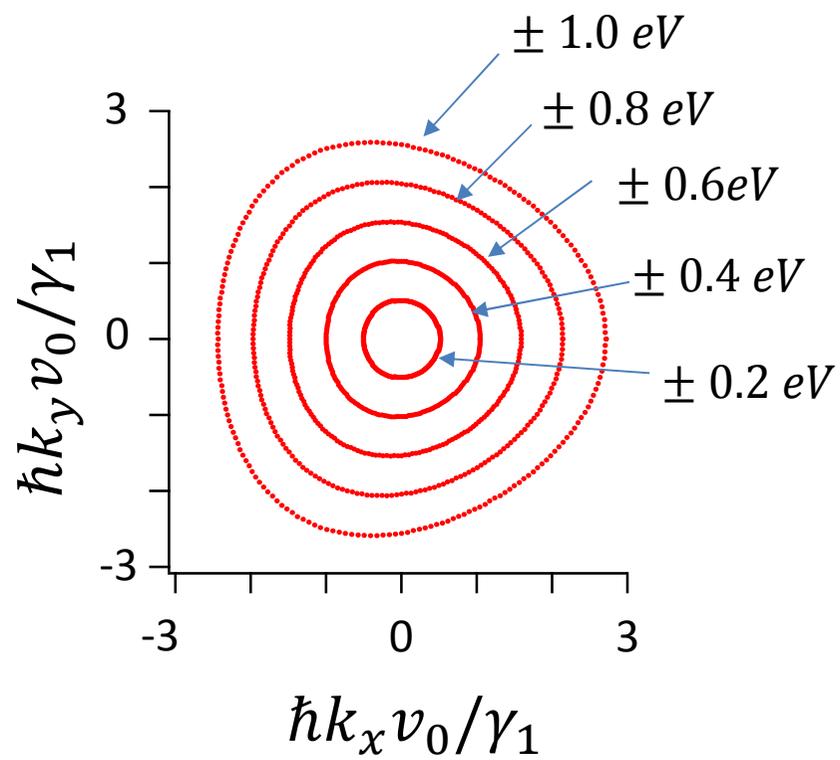
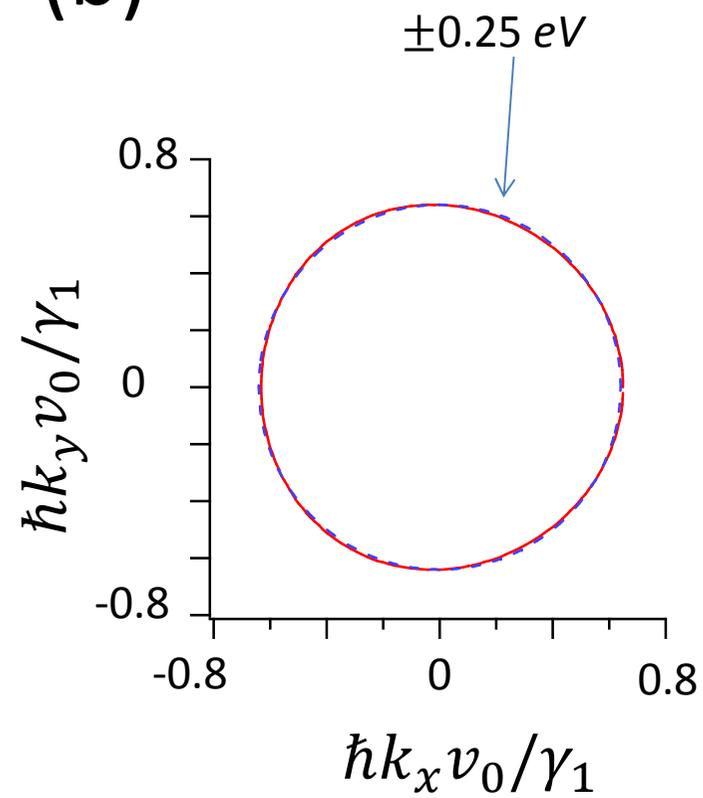

Fig. 8

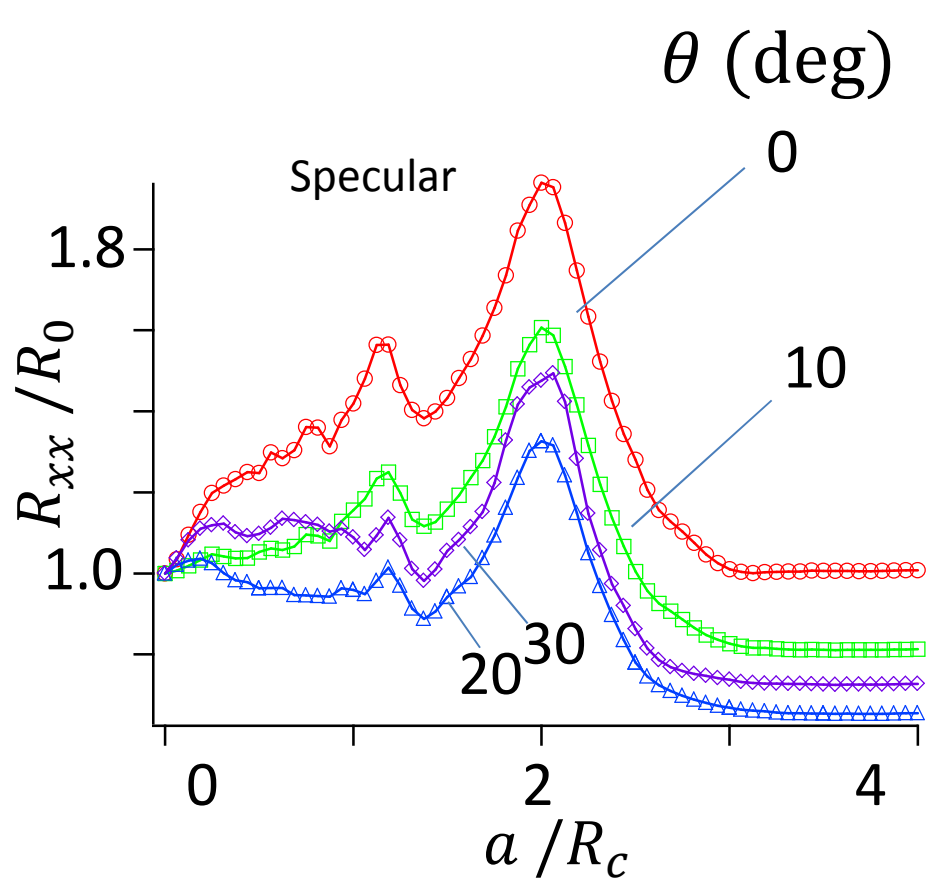
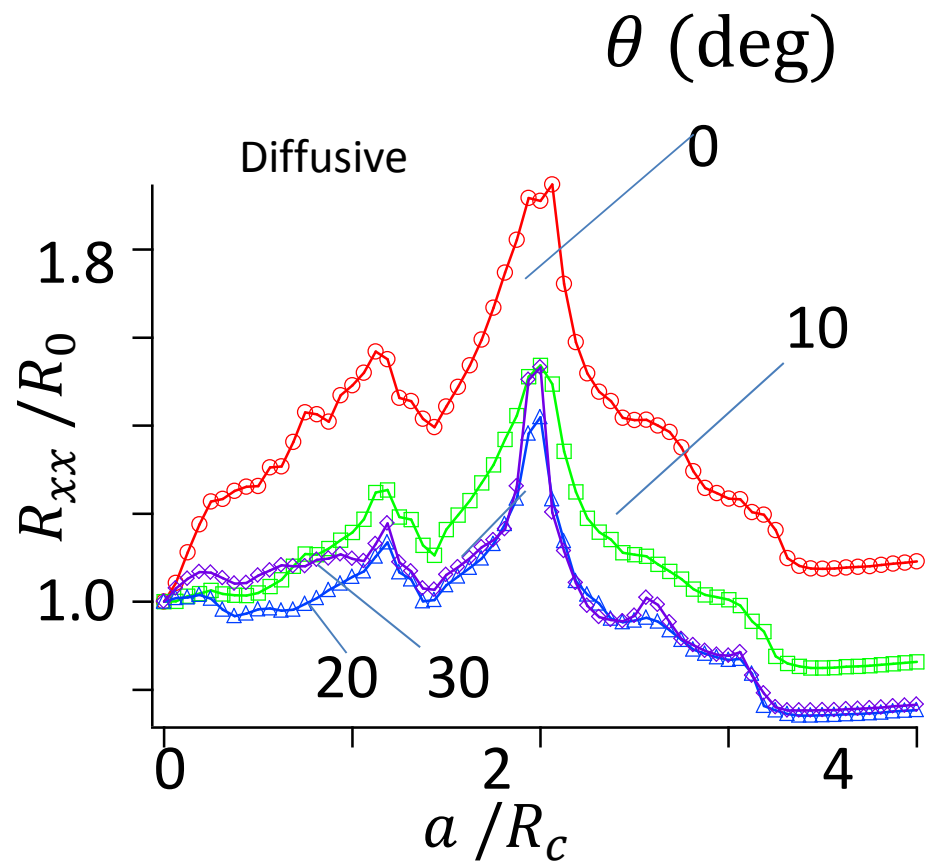